\documentclass[10pt,conference,letter,compsoc]{IEEEtran}

\usepackage{flushend}
\usepackage{paralist}
\usepackage[numbers]{natbib}
\usepackage{hyperref}
\usepackage{cleveref}
\usepackage[all]{hypcap}
\usepackage{graphicx}
\graphicspath{{pdfs/}}
\usepackage[normalem]{ulem}

\makeatletter
\newcommand\footnoteref[1]{\protected@xdef\@thefnmark{\ref{#1}}\@footnotemark}
\makeatother

\begin{document}
\title{Using Graph Properties to Speed-up GPU-based Graph Traversal: A Model-driven Approach}
\author{%
\IEEEauthorblockN{Merijn Verstraaten, Ana Lucia Varbanescu, Cees de Laat}\\
\IEEEauthorblockA{Informatics Institute, University of Amsterdam\\
Science Park 904\\
1098 XH Amsterdam\\
The Netherlands\\
\texttt{\{m.e.verstraaten, a.l.varbanescu, delaat\}@uva.nl}}
}

\maketitle
\begin{abstract}
While it is well-known and acknowledged that the performance of graph
algorithms is heavily dependent on the input data, there has been surprisingly
little research to quantify and predict the impact the graph structure has on
performance. Parallel graph algorithms, running on many-core systems such as
GPUs, are no exception: most research has focused on how to efficiently
implement and tune different graph operations on a specific GPU. However, the
performance impact of the input graph has only been taken into account
indirectly as a result of the graphs used to benchmark the system.

In this work, we present a case study investigating how to use the properties
of the input graph to improve the performance of the breadth-first search (BFS)
graph traversal. To do so, we first study the performance variation of 15
different BFS implementations across 248 graphs. Using this performance data,
we show that significant speed-up can be achieved by combining the best
implementation for each level of the traversal. To make use of this
data-dependent optimization, we must \emph{correctly predict} the relative
performance of algorithms per graph level, and \emph{enable dynamic switching}
to the optimal algorithm for each level at runtime.

We use the collected performance data to train a binary decision tree, to
enable high-accuracy predictions and fast switching. We demonstrate empirically
that our decision tree is both fast enough to allow dynamic switching between
implementations, without noticeable overhead, and accurate enough in its
prediction to enable significant BFS speedup. We conclude that our model-driven
approach (1) enables BFS to outperform state of the art GPU algorithms, and (2)
can be adapted for other BFS variants, other algorithms, or more specific
datasets.
\end{abstract}

\section{Introduction}
\label{sec:introduction}
Graph processing is an important part of data science, due to the flexibility
of graphs as models for highly interrelated data. Given the rapid growth of
dataset sizes, as well as the expected complexity increase of graph processing
applications, a lot of research focuses on parallel and distributed
solutions for graph processing~\cite{avery2011giraph,hong2015pgx,%
YGuo2015ccgrid,journal/VLDB/LuCYW14,journal/VLDB/HanDAOWJ14,%
conf/bigdata/ElserM13,conf/SIGMOD/SatishSPSPHSYD14,conf/IPDPS/GuoBVIMW14,%
graph500,beamer2013direction,Buluc16graph}.

With the increased popularity of graphics processing units (GPUs), some of this
research also ``migrated'' to these massively parallel architectures. Tempted
by the high-performance potential of GPUs, researchers investigate novel ways
to circumvent the (apparent) lack of regularity and data
locality~\cite{Lumsdaine2007} in graph processing to accommodate the massive
parallelism of GPUs. Therefore, several GPU-enabled graph processing frameworks
have emerged~\cite{YGuo2015ccgrid,wang2016gunrock,Heldens16IA3,%
khorasani2014cusha,becchi2013ipdps,gpuapsp:2008-15}.

Most graph processing frameworks, GPU-enabled or otherwise, simplify working
with graphs by hiding complexity; they maintain a separation between a
front-end that lets users specify their algorithm using high-level primitives
or domain specific languages, and a back-end that provides high-performance
implementations of these primitives for the given software or hardware
platform.

Unfortunately, there is no consensus which primitives form the ``canonical
set`` of operations required to implement graph algorithms. It is not even
clear if such a set exists. As a result, different frameworks choose
drastically different primitives to base their implementation on,
e.g.,~Gather-Apply-Scatter~\cite{gonzalez2012powergraph,xin2013graphx} or
vertex-centric
operations~\cite{malewicz2010pregel,avery2011giraph,McCune:2015:TLV}. To make
matters worse, there are often many different ways to implement the same
primitive, e.g.,~push versus pull for vertex-centric
primitive~\cite{beamer2013direction}.

If the performance of these primitives and implementations was only dependent
on the hardware, this would be a tractable optimisation problem: benchmark each
implementation for each piece of hardware and you can select a single ``best``
performing implementation for your system. This is an arduous task, but not
especially complicated.

But the performance of different implementations also depends on the structural
properties of the graph being processed. While it is accepted as common
knowledge that the performance  of data-dependent algorithms is impacted by the
structure of the data, little progress has been made, for graph processing, in
understanding how large this impact is, or in modeling the correlation between
the properties of the input data and the observed
performance~\cite{varbanescu15icpe,verstraaten2015quantifying,%
lehnert2016performance}.

Theoretically speaking, there are multiple solutions to understand and/or
quantify this correlation: (1) workload characterization and analytical
modeling, (2) controlled (micro)benchmarking, or (3) statistical modeling.

While workload characterization for graph processing on parallel systems has
been attempted, there is no such work to be found for GPUs. Moreover, there is
very little work in analytical modeling for parallel graph processing,
precisely because the strong dependence between hardware, algorithm, and
dataset.

In our previous work we attempted to determine the performance impact of
different graph properties by constructing an analytical model of the
sequential workload of the implementation. We attempted to link graph
properties to runtime behaviour~\cite{verstraaten2015quantifying}, via metrics
like:
\begin{inparaenum}[1\upshape)]
\item access coalescing,
\item occupancy,
\item branch divergence,
\item atomic retries.
\end{inparaenum}
Although the model was able to correctly estimate the amount of work per
(algorithm, dataset) pair, its generalization for parallel execution on GPUs
was less successful. Therefore, the resulting model had to be discarded due to
its low average accuracy.

Controlled (micro)benchmarking is another interesting approach, where by a
clever selection of the input datasets, we can study the changes in algorithmic
performance behavior and eventually isolate the performance impact of each
graph feature. In order for this approach to work, a large collection of
(synthetic) datasets is required, in which each property can be varied in
isolation.

No repository of such datasets exists. Instead, most research on graph
processing uses: 1) input data from several publicly available real world
datasets, such as SNAP~\cite{SNAP} and KONECT~\cite{KONECT}, or 2)
synthetically generated graphs using well-known generators/models, such as
R-MAT~\cite{chakrabarti2004r}, Kronecker graphs~\cite{leskovec2010kronecker},
and scale-free graphs~\cite{holme2002growing}.

The real world datasets are too noisy for systematic benchmarking; the graphs
vary in almost every property. The synthetic graphs, although more predictable
and controllable, still do not cover all properties of interest, and results
cannot be generalised to other types of graphs. We attempted to generate graphs
with the desired properties ourselves, but were unable to scale the generation
to graphs of sufficient size to do benchmarking~\cite{verstraaten2016}.

In this work, we focus on the third option towards understanding the
performance impact of dataset properties on graph processing: data-driven.
Specifically, we collect a lot of performance data from representative graphs,
and we use it as training data for a machine learning approach towards a model
that can predict performance for a given, unseen graph. While this approach
provides less insight into the actual correlations, it does provide a
systematic process for building a prediction model, and many tuning
possibilities in terms of features, variables, and actual methods.

Among all these methods, we initially selected Random Forests as the predictive
model, and found its accuracy to be good, but its applicability for fast
prediction too limiting. To simplify the model, we opted for a Binary Decision
Tree (BDT), and found it provides a good balance between accuracy and
applicability. Therefore, this work focuses on building and using a BDT-based
model to:
\begin{inparaenum}[1\upshape)]
\item determine which graph properties have the biggest impact on algorithm
    performance, and
\item improve the BFS performance by predicting and deploying the best
    performing algorithm for each BFS level.
\end{inparaenum}

We have applied this model on the 248 graphs from KONECT; for each graph we
used 20 different traversals, starting from 20 different nodes, and collected
the features and performance indicators for each traversed level. We trained
the model on a 70\% random split of the data. When testing on the remaining
30\% of the data, the model predicts with 96\% accuracy. Finally, we use the
model as a switch predictor for a level-switching adaptive BFS. With this
adaptive BFS, we outperform two popular graph processing systems for GPUs: we
gain on average $2{\times}$ over Gunrock~\cite{wang2016gunrock} and
$15{\times}$ over LonestarGPU~\cite{burtscher2012quantitative}.

The main contributions of this paper are the following:
\begin{itemize}
\item We show that the performance of different BFS implementations varies
    dramatically across both graphs and iterations within the same graph, with
    differences of up to two orders of magnitude
    (\Cref{sec:experiments}).

\item We create a binary decision tree model that can predict which BFS
    implementation to use at every level of BFS.
    (\Cref{ssec:model_building}).

\item We show that our decision-tree model is accurate enough and fast enough
    to evaluate online, allowing for dynamic switching of implementations
    during BFS (\Cref{ssec:feasibility}).

\item We demonstrate that using our dynamic switching BFS approach, enabled by
    the accurate prediction of our model, may lead to performance gains of
    ${\sim}20\%$ to ${\sim}80\%$ over the best single-algorithm BFS
    (\Cref{ssec:feasibility}).
\end{itemize}

\section{Background}
\label{sec:background}
For readers unfamiliar with binary decision trees or GPGPU processing, this
section provides some basic information required to understand the rest of this
paper.

\subsection{Decision Trees}
\label{sec:dec-trees}
Decision trees are a non-parametric, supervised learning
technique~\cite{breiman1984classification}. They come in two flavours,
\emph{classifiers} and \emph{regressors}. After constructing a decision tree
from a \emph{training set} of $(X, Y)$ pairs, where $X$ is a tuple of 1 or more
inputs and $Y$ is a tuple of 1 or more outputs, we can use this tree to predict
the expected output tuple for a given input tuple. The working assumption is
that the original learning set is representative of the observable input-output
pairs.

To construct a (binary) decision tree we recursively partition the learning set
along one of the $N$ input parameters, preferring the parameter (and value)
that has the strongest discriminating power, i.e.~the one that produces the
partitioning closest to 50--50. The split effectively means that we assign all
elements smaller than the chosen value to the left branch and the others to the
right branch. This process repeats until we reach a stopping criteria --- e.g.,
the maximum tree height, minimal bucket size.

After the stop condition is hit, the output for each bucket is computed. If the
bucket contains 1 output value, or multiple equal outputs, this is the result
for that bucket. If there are multiple unequal values, the outputs are
normalised to a single output. For regressions this usually done by averaging
all values in the bucket. For classification, this is usually by selecting the
``most likely'' value in the bucket, although more complicated strategies
exist.

We can now use this computed binary tree as a prediction model. To compute a
prediction using this tree for an input tuple, we compare the tuple's value
against the parameter stored in the tree at each level, and walk down the
correct branch. Once we reach a leaf node, we return the value computed above
for that bucket as resulting prediction.

Different decision tree algorithms use different criteria to compute which
parameter and value to split the tree on. We used the implementation in Python
library scikit-learn~\cite{scikit-learn}. This uses an optimised algorithm
based on the CART~\cite{breiman1984classification} algorithm. This algorithm
splits the training set based on which parameter produces the large reduction
in Gini impurity. Gini impurity is a measure of how often an element in a
subset would labelled wrong if all elements in the subset were labelled
randomly, according to the distribution of labels in that subset.

We can estimate the importance of each input parameter by computing its Gini
importance. This is done by computing the total, normalised, reduction of Gini
impurity resulting from that parameter. Similar importance measures exist for
other decision tree algorithms, meaning that we can relatively easily compute
the importance of each parameter in our input.

To summarise the advantages of decision trees:
\begin{inparaenum}[]
\item They are simple to understand and interpret;
\item Small trees can be visualised;
\item They require little to no data preparation;
\item Prediction cost is logarithmic in the number of data points used;
\item They can handle both categorical and numerical data;
\item They can handle multi-output problems;
\item Parameter importance is known after training.
\end{inparaenum}

The main downsides of decision trees are:
\begin{inparaenum}[]
\item They are prone to overfitting;
\item Small differences in data can result in drastically different results
    (i.e., unstable models);
\item Constructing optimal decision trees is NP-complete under several aspects
    of optimality;
\item They cannot represent all concepts easily (XOR, parity, multiplexer
    problems);
\item Biased trees are easily created, if some classes dominate.
\end{inparaenum}

Due to the way trees are constructed, overfitting issues can become more
pronounced if the input parameters in the learning set are not uniformly
distributed across the range we intend to predict against. Additionally, as the
number of input parameters increases it becomes exponentially more costly to
compute the best discriminator, which in turn makes the algorithm slower and
increases the risk of bias and overfitting. To reduce and detect overfitting we
train our decision trees on a random subset of our data points and
cross-validate our model against the unseen data points. See
\Cref{ssec:overfitting} for a more detailed discussion on concerns related to
overfitting.

\subsection{Graph Processing}
Graphs are collections of entities (called \emph{nodes} or \emph{vertices}) and
relationships between them (called \emph{edges}) --- $G=(V,E)$. Graph
processing typically implies some transformation of the original graph by
traversing its edges and visiting is nodes. The simplest example itself being a
traversal itself, where, given a starting node (also called the root node), the
algorithm has to visit all the nodes accessible from the root, eventually
providing shortest path between the root and all accessible nodes. There are
two types of traversals: the Breadth-First Search (BFS) and Depth-First Search
(DFS). In this work we focus mainly on BFS.

In general, graph processing applications --- and traversal is a good example
--- are not easy to parallelize due to their properties: low
compute-to-communication ratio, data-dependent behavior, low data locality,
variable parallelism, and load imbalance, etc. Thus proposing efficient
parallel algorithms for these algorithms is a challenge.

\subsection{General Purpose GPU Programming}
GPUs (Graphics Processing Units) are the predominant accelerators for high
performance computing. A GPU is, currently, a very good example of a many-core
architecture: it has hundreds to thousands of slim cores, grouped into
streaming multi-processors with local shared memory and/or caches; it provides
a hierarchical memory model, with large register files per multi-processor,
local L1 and shared L2 caches, and a global memory which increases in size with
every generation.

GPUs promise huge theoretical performance: the peak performance of a regular
card can easily be in excess of 2 TFLOPs computational throughput and 200--300
GiB/s memory bandwidth. With such performance numbers, sooner or later, all
computational domains will investigate whether GPUs are a suitable target for
their computational needs.

Graph processing is no exception: our work focuses on understanding the
potential for GPUs to boost the performance of graph processing algorithms,
which are notoriously difficult to parallelize efficiently. In this work, we
use NVIDIA GPUs, due to their superior programmability provided by
CUDA\footnote{CUDA is the native programming model for NVIDIA GPUs; it is
proprietary to NVIDIA, but has a huge ecosystem of libraries and helpful
tools, unmatched by models like OpenCL or OpenACC.}.

The idea behind the programming model is simple: CUDA provides a mapping of the
programming model concepts onto the hardware, while preserving a sequential
programming model per thread. For the actual computation, programmers focus on
implementing the single-thread code, called a \emph{kernel}; they further write
the \emph{host code} to launch enough threads to (1) cover the space of the
problem, and (2) provide enough potential for latency
hiding~\cite{volkov2016understanding,nvidiacuda}. The threads that execute the
kernel are grouped into \emph{thread blocks}, which are scheduled on the
streaming multi-processors. All blocks form \emph{a grid}, which effectively
contains all the logical threads that are to be scheduled and eventually
executed on the cores themselves.

In terms of the execution model, NVIDIA GPUs work with \emph{warps}. A
\emph{warp} is a group of 32 threads that work in lock-step: they all execute
the same instruction on multiple data. This model is called SIMT --- Single
Instruction Multiple Thread --- and enables high performance by massive
parallelism, but is unable to handle diverging threads, it also poses
programming challenges to avoid the severe penalties that any inner-warp load
imbalance might bring. Besides thread divergence, other performance challenges
in GPU programming are the abuse of atomic operations and lack of
\emph{coalescing} for the main memory accesses.

Our software stack is based on C++ and CUDA, and it is available
on~\href{https://github.com/merijn/gpu-benchmarks}{GitHub}%
\footnote{\url{https://github.com/merijn/gpu-benchmarks}}.

\section{Experiments}
\label{sec:experiments}
We benchmarked our 15 different BFS implementations on the graphs found in the
KONECT~\cite{KONECT} dataset, measuring both the total time and the time taken
for each level of BFS. We used these results to train and validate our Binary
Decision Tree (BDT) model.

\subsection{BFS Implementations}
We wrote 5 different implementations of BFS, and for each of these 5
implemented 3 variants. These 5 implementations consist of: 2 edge-centric
implementations (edge list and reverse edge list), 2 vertex-centric
implementations (vertex push and vertex pull), and 1 virtual warp-based
implementation based on the work by \citeauthor{Hong2011}~\cite{Hong2011}. The
3 different variants are based on how the new frontier size is computed at the
end of each BFS level. In this subsection we describe how these versions differ
from each other.

Each algorithm starts by initialising all depths to infinity, then initialising
the root node's depth to 0. During every level of BFS we compute the frontier
size, that is, the number of vertices that have been assigned a new depth.

\subsubsection{Edge List \& Reverse Edge List}
For every level of BFS these edge-centric implementations launch \emph{one CUDA
thread per edge}. If the depth of the origin vertex is equal to the current BFS
level, then the depth of the destination vertex is updated to the minimum of
its current depth and the BFS level plus one.

The edge list implementation uses the outgoing edges of every vertex, whereas
the reverse edge list implementation use the incoming edges of every vertex.
This difference affects the amount of memory coalescing and the access patterns
exhibited at runtime.

The advantage of these edge-centric parallelisations is that they never suffer
from workload imbalance, every thread in a warp performs the same amount of
work. The fact that many threads have to read the depth of the same origin
vertex helps with coalescing memory access. The downside is that they result in
many parallel updates, resulting in many heavily contested atomic updates.

\subsubsection{Vertex Push \& Vertex Pull}
For every level of BFS these vertex-centric implementations launch one CUDA
thread per vertex. For the push implementation, if the vertex its depth is
equal to the current BFS level, the thread iterates over all its neighbours,
updating their depth to the minimum of their current depths and the BFS level
plus one. For the pull implementation, if the vertex has no depth yet, the
thread iterates over its neighbours until it encounters one whose depth matches
the current BFS level, if this happens it sets its depth to the current BFS
level plus one.

Both implementations are susceptible to workload imbalance, if vertices with
wildly different degrees are in the same warp, this divergence will result in
reduced performance. The push version, similar to edge-centric implementations
generates a lot of concurrent updates, requiring a large number of atomic
operations. However, if the frontier is small it avoids many useless reads,
since the depth of every vertex is only read once.

The pull version does not require atomics as the depth of a vertex is only ever
touched by one thread. The downside is that, if none of a vertex its neighbours
are in the frontier, a lot of time is wasted iterating over neighbours for
nothing. As such pull becomes more efficient as more vertices are in the
frontier, since a thread can stop looping over its neighbours as soon as it
discovers one in the frontier.

\subsubsection{Vertex Push Warp}
As mentioned above, this implementation is based on the vertex push
implementation, but rather than assigning one thread per vertex, it uses the
virtual warp method described in \cite{Hong2011}, which attempts to mitigate
the negative impact produced by workload imbalance between threads.

The basic principle is the same as with vertex push, but rather than assigning
a single thread per vertex, we divide the warps into smaller ``virtual warps''.
Each virtual warp gets assigned a number of vertices, equal to its number of
threads. However, instead of each thread processing the edges
for one vertex, all $N$ threads process the edges of the first vertex in
parallel, then the edges of the second vertex are processed in parallel, and so
on until all vertices assigned to the virtual warp have been processed.

This reduces the amount of load imbalance occurring within a virtual warp,
since the workload of a virtual warp is spread out equally across that virtual
warp. However, the optimal size of the virtual warp is challenging to
determine. Moreover, the different graphs and even different levels of the
graph also require tuning of the warp sizes for best performance.

\subsubsection{Variants}
In every BFS level, 0 or more new vertices get discovered. These vertices will
form the frontier for the next level. We need to track the size of the
frontier, since the algorithm terminates when no new nodes are discovered. To
accomplish this task, each thread tracks how many new vertices it has
discovered. At the end of each BFS level, we need to aggregate these counts to
compute the new frontier size.

We implemented three different methods to do this aggregation. The first
variant simply uses a global counter and every thread performs an atomic
addition on this counter. This approach results in heavily contested atomics
operations. Most of the CUDA literature suggests that we can reduce the
contention and number of atomic operations by first performing a reduction
within a warp or block~\cite{harris2007optimizing}, before performing the
global atomic operation. Thus, the second variant performs a warp reduction
before the global atomic update, while the third variant performs a
warp-and-block reduction before atomically updating the frontier size.

\subsection{Experimental Setup}
All measurements have been performed on an NVIDIA TitanX, using version 7.5 of
the CUDA toolkit. The source code of these benchmarks can be found on GitHub%
\footnote{\url{https://github.com/merijn/GPU-benchmarks}}.

As for datasets, we retrieved all the graphs from the KONECT repository and ran
each of the $3 \times 5$ implementations described above on all of them.
Additionally, for every graph, we used 20 different root vertices\footnote{For
the graphs with less than 20 vertices we used every vertex as root.}. All
results presented here are averaged over 30 runs, and exclude input data and
result transfer times.

The first observation we made about our results is that both the warp reduction
and warp-and-block reduction variants perform significantly worse than the
direct atomic versions, in all cases. Therefore, we do not include any of these
variants in the plots of this paper, to keep the results easily readable.

\begin{figure*}[tbh]
\includegraphics[width=\textwidth]{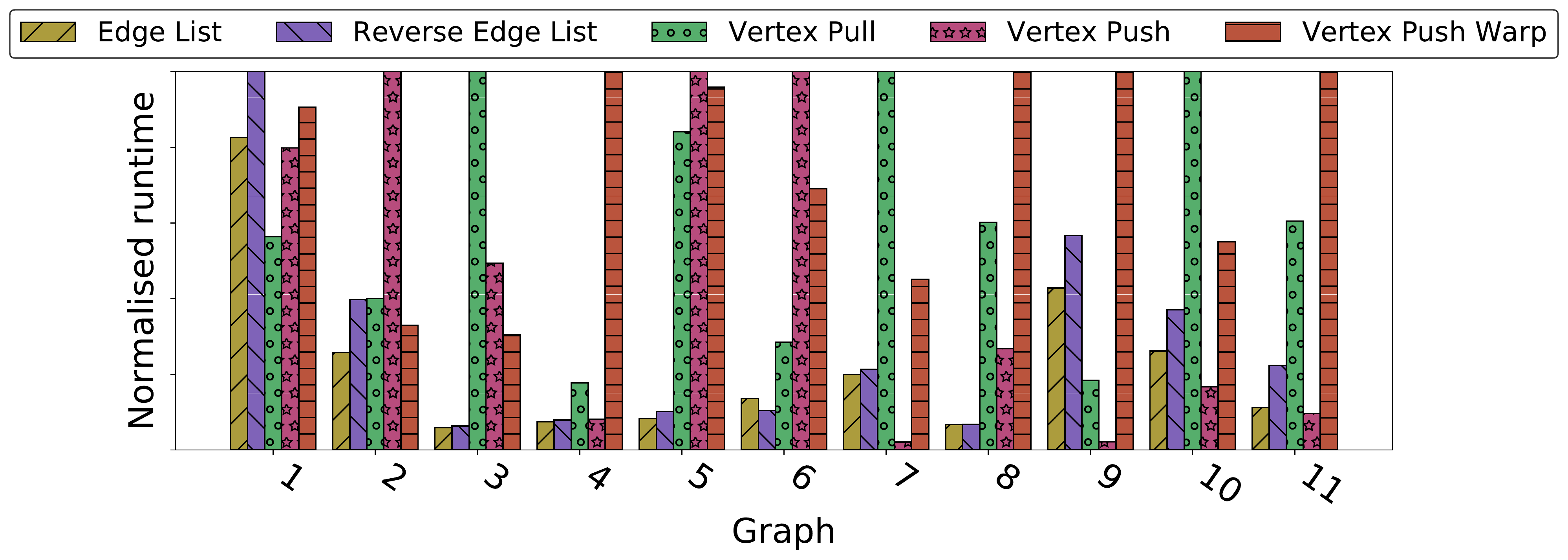}
\caption{Normalised total runtimes for different BFS implementations on KONECT
graphs. See \Cref{tab:graphs} for the details of the input graphs.}
\label{fig:bfstotals}
\end{figure*}

\Cref{fig:bfstotals} shows the runtimes, normalised to the slowest
implementation for each graph, for a selection of KONECT
graphs\footnote{\label{foot1}The full set of performance plots is available
at:\\ \url{https://staff.fnwi.uva.nl/m.e.verstraaten/}\\(These will be moved to
more permanent hosting for the camera-ready version.)}.

\begin{table}
\centering
\begin{tabular}{llrrl}
    No. & Graph & \# Vertices & \# Edges\\
  \hline
    1 & actor-collaboration & 382,219 & 30,076,166 \\
    2 & ca-cit-HepPh & 28,093 & 6,296,894 \\
    3 & cfinder-google & 15,763 & 171,206 \\
    4 & dbpedia-starring & 157,183 & 562,792 \\
    5 & discogs\_affiliation & 2,025,594 & 10,604,552 \\
    6 & opsahl-ucsocial & 1,899 & 20,296 \\
    7 & prosper-loans & 89,269 & 3,330,225 \\
    8 & web-NotreDame & 325,729 & 1,497,134 \\
    9 & wikipedia\_link\_en & 12,150,976 & 378,142,420 \\
   10 & wikipedia\_link\_fr & 3,023,165 & 102,382,410 \\
   11 & zhishi-hudong-internallink & 1,984,484 & 14,869,484 \\
  \hline
\end{tabular}
\caption{Details for the input graphs shown in \Cref{fig:bfstotals} and
\Cref{fig:bfscomparison}.}
\label{tab:graphs}
\end{table}

The plots in \cref{fig:bfstotals} were selected to illustrate the point we made
in the introduction: the performance of different implementations can vary by
orders of magnitude across input graphs.This effectively means that when
(accidentally) choosing the worst algorithm, one can loose 1--2 orders of
magnitude in performance for a BFS traversal compared with the best option. In
turn, this means an informed decision about the algorithm to be used for a
given graph is not only desirable, but quite important for any efficiency
metric. But this is no easy task: no models are available to determine the best
or the worst algorithm for a given graph traversal task.

%

One of the reasons for which predicting the best algorithm for the entire graph
is difficult is the huge performance difference that can be observed during
traversing a single graph: (1) between two different BFS levels in the same
graph, the performance of the same algorithm can vary up to an order of
magnitude, and (2) per-level, the differences between different algorithms can
be up to four orders of magnitude.

In over 75\% of the cases the difference between the best and worst
implementation for a single level is more than 2 orders of magnitude, and 4
orders of magnitude in the worst cases. This, combined with the fact that each
algorithm wins at least once and loses several times, makes a random choice far
from ideal for determining the implementation to use at a given level.

\begin{figure*}[tbh]
\includegraphics[width=\textwidth]{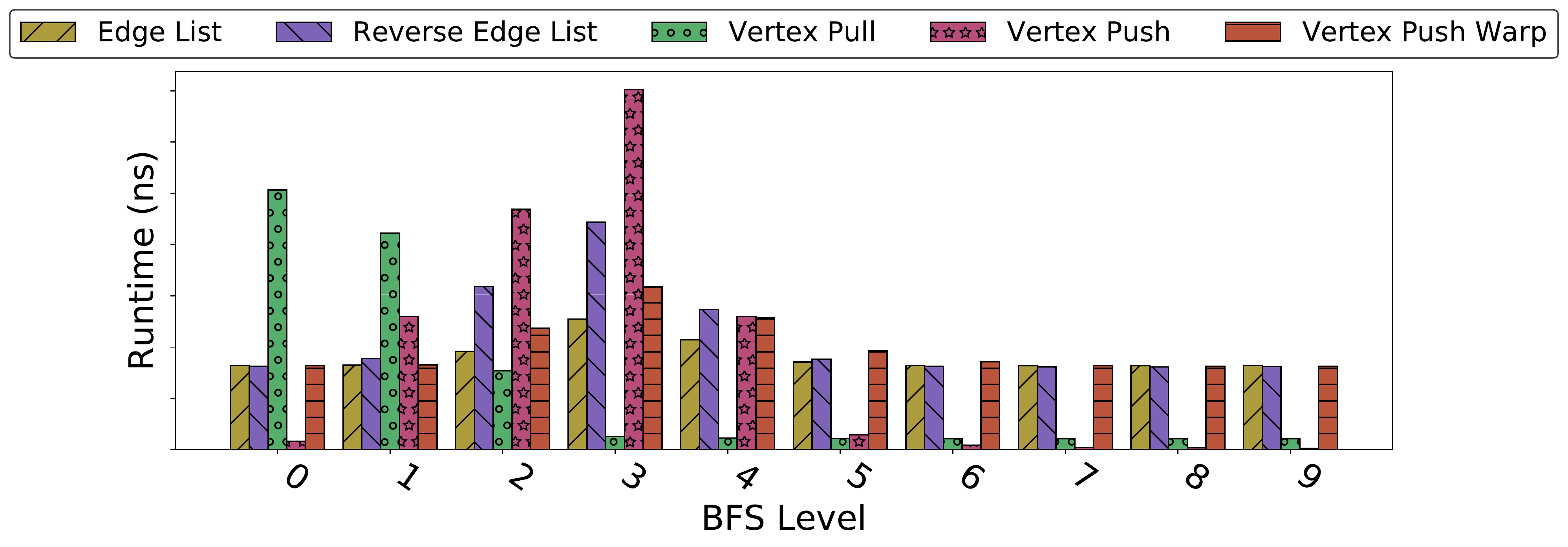}
\caption{Runtimes of different BFS implementations per level of the
\texttt{actor-collaborations} graph from KONECT.}
\label{fig:bfslevels}
\end{figure*}

To illustrate these large performance gaps, \cref{fig:bfslevels} presents an
example of the performance of the five main BFS implementations, per-level, for
the \texttt{actor-collaborations} graph\footnoteref{foot1}. We see that the
vertex push implementation dominates on most levels, but on a handful of levels
it performs so badly that its overall performance becomes terrible. Such
behaviour is a strong indication that switching algorithms at every level might
be even better than devising a model to detect the best overall solution.

\section{Modeling BFS Performance}
\label{sec:modeling}
There is no consensus on which of a graph's structural properties impact the
performance of graph algorithms. Our previous modeling attempts
in~\cite{verstraaten2015quantifying}, combined with our experiences while
optimising and developing our implementations lead us to believe that the graph
size and degree distribution are the biggest factors when it comes to neighbour
iteration. Additionally, the work of \citeauthor{beamer2013direction} and
\citeauthor{becchi2013ipdps} on adaptive
BFS~\cite{beamer2013direction,becchi2013ipdps}, and the observed runtime
changes across levels, indicates that the behaviour at each level is dependent
on the size of the frontier discovered in the previous level, and the
percentage of the graph that has already been discovered. Therefore, these are
the features we focus on when building a performance prediction model.

Further in this section we describe the training process used to build our
model, discuss its accuracy and the applicability for online performance
prediction, and analyse the feasibility of a dynamically switching BFS.

\subsection{Building the model}
\label{ssec:model_building}
The experiments described in the previous section provide us with timing data
for all 15 of our algorithms and each BFS level of all KONECT graphs. This data
allows us to compute the fastest implementation for each BFS level and graph
combination. We built a training set where we associated every measured BFS
level with the structural properties of the graph it was run on, and the
level's specific information. We consider the following relevant features for
our model:
\begin{description}
\item[\textbf{Graph size:}]\hfill\\
the number of vertices and edges in the graph.
\item[\textbf{Frontier size:}]\hfill\\
either as absolute number of vertices or as percentage of the graph's vertices.
\item[\textbf{Discovered vertex count:}]\hfill\\
either as absolute number of vertices or as percentage of the graph's vertices.
\item[\textbf{Degree distribution:}]\hfill\\
represented by the 5 number summary and standard deviation of in, out, or
absolute degrees.
\end{description}

The models described in the rest of this section consist of binary decision
trees trained to predict the best performing algorithm for a given level of
BFS, based on a mix of the above properties. We remind the reader that the
predicted value for each leaf in the tree is based on a majority vote on what
is ``most likely'' given the values in the bucket associated with that leaf ---
see \cref{sec:dec-trees}). However, there can be special cases when the model
cannot deliver a prediction because no single value can be computed based on
the values in the bucket --- e.g., as a result of a tie.

In such cases we talk about an ``unknown prediction''. We opted to resolve
unknown predictions for level $N$ of a BFS by repeating the prediction for
level $N-1$. In case of an unknown prediction at the first level of BFS, we
default to predicting the edge list implementation, as this implementation is
the least likely to have extremely bad performance, which should reduce the
likelihood of early unknown predictions resulting in significant performance
loss.

\subsection{Model Accuracy}
\label{ssec:accuracy}
We define the optimal BFS traversal of a graph as the traversal where the
fastest of our implementations is used at every level. To evaluate the accuracy
of our models, we take this optimal runtime as a reference (i.e., as 1) and
evaluate the predicted and observed runtimes as a slowdown compared to this
theoretical optimum. The larger the gap, the further away we are from the
optimal performance.

Of the models we have trained using our experimental dataset, the one that
performs best for this prediction task is one based on the following four
features: \emph{graph size}, \emph{percentage of vertices discovered},
\emph{the distribution of out degrees}, and \emph{the number of vertices in the
current frontier}. In \cref{tab:accuracy} we compare the model's predictions
and the different implementations against the optimal runtime across all KONECT
graphs. The optimal runtime is the execution time of the optimal BFS traversal.
The ``oracle'' runtime shows the fastest algorithm for every graph when dynamic
runtime switching during the BFS computation is disabled (i.e., the best
performing non-switching algorithm for each graph).

\begin{table*}
\centering
\begin{tabular}{lrrrrrr}
  Algorithm & Optimal & 1--2$\times$ & ${>} 5 {\times}$ & ${>} 20 {\times}$ & Average & Worst\\
  \hline
  Predicted & $56\%$ & $41\%$ & $1\%$ & $0.5\%$ & $1.40{\times}$ & $236{\times}$\\
  Oracle & $23\%$ & $55\%$ & $2\%$ & $0\%$ & $1.65{\times}$ & $8.5{\times}$\\
  Edge List & $10\%$ & $61\%$ & $7\%$ & $0.4\%$ & $2.22{\times}$ & $38{\times}$\\
  Rev. Edge List & $5\%$ & $59\%$ & $15\%$ & $0.6\%$ & $2.92{\times}$ & $50{\times}$\\
  Vertex Pull & $0\%$ & $15\%$ & $27\%$ & $24\%$ & $38.62{\times}$ & $2,671{\times}$\\
  Vertex Push & $9\%$ & $15\%$ & $53\%$ & $29\%$ & $39.66{\times}$ & $1,048{\times}$\\
  Vertex Push Warp & $0\%$ & $0\%$ & $3\%$ & $30\%$ & $18.69{\times}$ & $97{\times}$\\
  \hline
\end{tabular}
\caption{Algorithm performance compared to theoretical optimum over all the
graphs in KONECT.}
\label{tab:accuracy}
\end{table*}

Following our model's predictions, we observe an average runtime of $1.40$ of
optimal, which effectively means a 40\% slowdown compared to optimal. However,
the oracle average runtime is $1.65$, or 65\% slowdown compared to optimal. In
other words, by following the model's prediction, we can obtain a 15\% speedup
compared to this oracle. In practice, the potential gain is more significant,
as such an oracle does not exist.

\begin{figure*}[tbh]
\includegraphics[width=\textwidth]{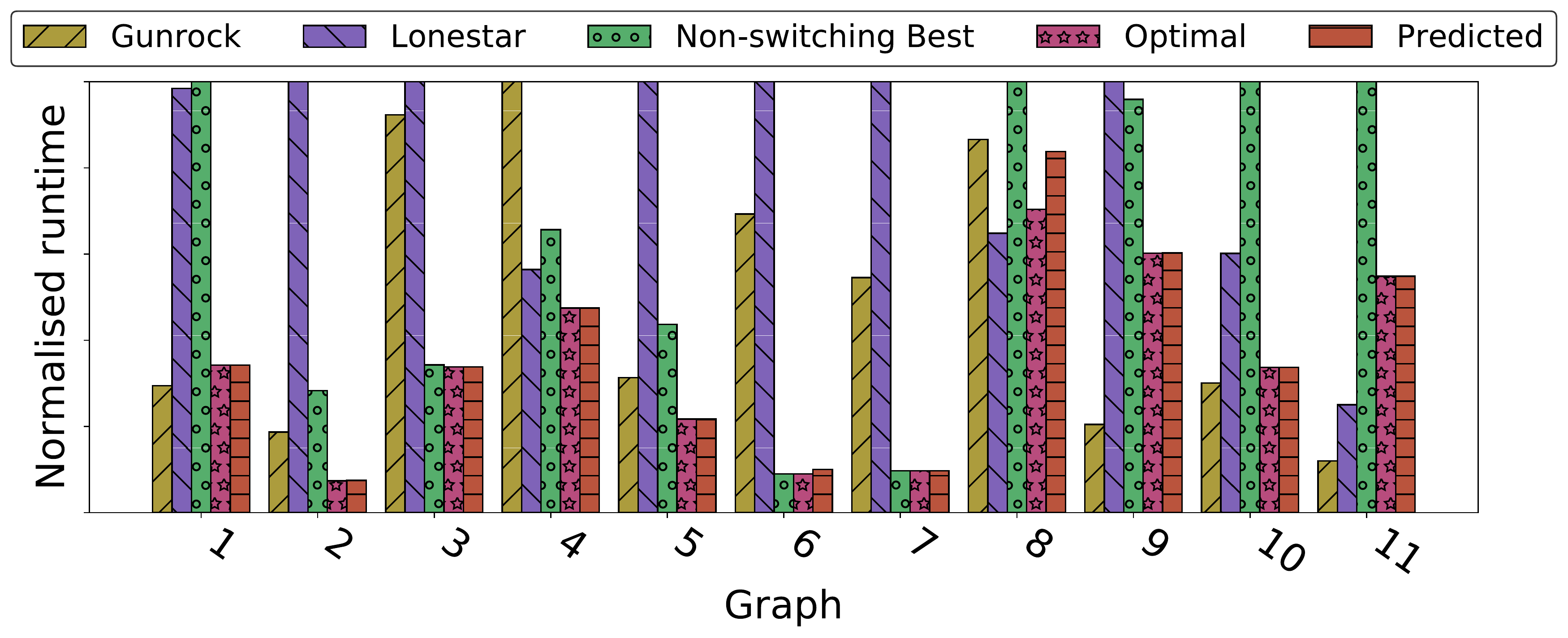}
\caption{Comparison of normalised runtimes of different BFS implementations,
predicted performance, and existing optimised BFS implementations on KONECT
graphs. See \Cref{tab:graphs} for the details of the input graphs.}
\label{fig:bfscomparison}
\end{figure*}

These results show that our model results in considerable speedup compared to
our individual implementations. However, speedup results are only as good as
the baseline. We compare against two existing GPU graph processing frameworks
to establish how much ``real world'' performance we gain by using the model
described in this paper.

\Cref{fig:bfscomparison} compares our results against the state-of-the-art GPU
graph processing framework Gunrock~\cite{wang2016gunrock} and the slightly
older BFS benchmark LonestarGPU~\cite{burtscher2012quantitative}, across a
selection of KONECT graphs. We benchmarked both Gunrock and LonestarGPU on the
same hardware, using 148 different KONECT graphs. On average, Gunrock achieves
a performance of $2.9{\times}$ of our theoretical optimum. LonestarGPU manages
$21{\times}$ of optimal. Our model's $1.4{\times}$ of optimal means that we
are, on average, $2{\times}$ faster than Gunrock.

\subsection{Feasibility of Dynamic Switching}
\label{ssec:feasibility}
The results presented in~\Cref{sec:experiments} indicate a significant
performance improvement when switching between different implementations at
runtime, but these gains can only be realised if the cost of switching does not
outweigh the gains. Thus, the feasibility of dynamic switching effectively
depends on (1) how long it takes to compute a prediction based on the trained
model, and (2) how expensive a ``context switch'' is between these
implementations.

To determine whether the runtime prediction is cheap enough, time-wise, we
extracted the binary decision tree from scikit-learn, converted it into a C
array-based data structure, and corresponding lookup function. We then measured
the time it takes to compute a prediction for each graph and level in our
dataset. The average prediction time is 144~ns, with a standard deviation of
161~ns, and a maximum of 16~$\mu$s. For comparison, the average BFS level
computation in our dataset takes 28~ms. Thus, computing the prediction is, on
average, insignificant compared with the actual processing of each level.

We still need to evaluate how expensive the ``context switch'' between
implementations is. We note that most of our implementations operate on
different representations of the graph, so to switch to a different
implementation, we need to (a) either generate/bring the new representation in
memory, or (b) simply keep all representations in memory.

Option (a) is not really feasible, because transferring data to and from the
GPU is generally slow, and doing so for each level would become prohibitive,
performance-wise. For option (b), we mush combine the two different
representations into one, which is is a feasible solution, a classical
time--space trade-off, where we trade memory for faster computation.

The two main graph representations we use are a Compressed Sparse Row (CSR) for
the vertex-centric implementations, and an edge list for the edge-centric
implementations. We can combine the two by simply storing the origin vertex for
every edge in our CSR. This increases the storage from 1~int/vertex and
1~int/edge (for CSR) and 2~int/edge (for edge list), to 1~int/vertex and
2~int/edge. This is not very expensive, memory-wise: it is a mere 38~MiB for a
graph of 10,000,000 edges.

\subsection{Overfitting \& Generality Concerns}
\label{ssec:overfitting}
As mentioned in \cref{sec:dec-trees} we took the standard precaution of
training our model against a subset of 70\% of our data and validating its
accuracy against a separate test set of 30\% of the data points. In this
validation, the model accurately predicts the fastest algorithm in 95\% of the
cases. The average difference in runtime between the fastest implementation and
our predicted implementation is 9.9\%.

From this data we can conclude that the model's accuracy is high with regards
to our KONECT repository of graphs. However, we expect the portability of the
model to be highly correlated with how representative the training set is for
the test set. For example, if we train the model on social networks graphs
only, we expect it to be better at predicting the best BFS for social networks,
and less so for, say, road networks. Therefore, we recommend that the actual
training and modeling process is driven by the prediction goals.

For example, If the goal is to build a generic model to predict most graphs,
using a large variety of graphs for training is mandatory. A collection like
KONECT is a good start, but we have two indicators that it is not a balanced
collection.

First, while validating the model against the KONECT data set, we noticed that
bad model predictions are correlated with several classes of graphs, such as
bipartite graphs, and graphs with extremely skewed degree distributions, which
are less represented in the repository (and, thus, in our training data).

Second, we also used the model to predict algorithm selection for 19 graphs
from the SNAP repository. For these graphs, our model performed significantly
worse than for KONECT, achieving an average runtime of $3{\times}$ compared to
optimal, or $2{\times}$ worse compared to an oracle that predicts the best
non-switching algorithm. There are two plausible causes for these
mis-predictions: (1) KONECT is indeed not representative enough of all graphs,
causing our model to miss due to a lack of samples for specific cases, and (2)
there are important structural properties not included in our current training
data.

On the other hand, if the goal is to have a model tweaked for a specific type
of graphs --- e.g., social or road networks --- only a subset of the graphs in
public repositories can be useful for training. Whether there are sufficient
such graphs depends on many factors. However, this analysis deserves a
dedicated study, focused on determining what is the ideal size and composition
of a specialized training set; such a study is beyond the scope of this work.

To summarize, we make no portability claims or guarantees of the trained model
for more specialized repositories, and we recognize the limitations of our
training dataset. However, the training and prediction processes are both
straightforward and generic, and can be easily applied for different training,
eventually improving/tuning the predictor to match the goal.

\section{Related Work}
\label{sec:related}
This section briefly introduces relevant work from the three research
directions closest to this work: (1) the design and implementation of parallel
graph processing algorithms, (2) graph processing frameworks and systems for
graph processing, and (3) the use of machine learning for performance modeling
and prediction.

\subsection{Algorithms}
Despite the advances in large-scale graph traversal algorithms, like direction
switching BFS~\cite{beamer2013direction}, distributed-memory
BFS~\cite{Buluc16graph}, and the matrix-based graph processing
solution~\cite{sparsechapter}, there's still no single best BFS traversal
solution. This is because BFS is highly dependent on the graph properties, with
different algorithms and/or implementation eventually suffering from different
bottlenecks.

When combining this with complex, massive parallel machines like the
GPUs~\cite{gpuapsp:2008-15}, the performance gaps are even more difficult to
predict. In our work, we steer away from attempting to devise yet another
algorithm for BFS, and focus on using the best existing solution in each
iteration. A somewhat similar approach has been attempted
in~\cite{becchi2013ipdps}, but our solution combines more algorithms and uses a
more deterministic, systematic switching criterion. Moreover, our approach can
be extended to incorporate additional BFS versions, as long as sufficient
performance data are available for training.

\subsection{Graph Processing Systems}
The new challenges of graph processing have also reflected in the amount of
systems and frameworks designed for efficient, high performance graph
processing~\cite{McCune:2015:TLV,Doekemeijer14survey}. From these systems, a
handful of GPU-enabled systems have also
emerged~\cite{wang2016gunrock,burtscher2012quantitative}, combining clever BFS
algorithms with specific GPU-based optimisations~\cite{Hong2011}.

Still, none of them can claim the absolute best performance for the same
reason: the diversity of graphs and their properties lead to high performance
variability for all these systems~\cite{guo2015empirical}. This work is
complementary to such systems: our switching approach can be, in principle,
incorporated in these frameworks. Performance-wise, we are competitive against
such systems (see \Cref{ssec:accuracy}), thus exploring the potential of
incorporating such an adaptive approach into an existing system is promising as
future work.

\subsection{Machine Learning for Performance Modeling}
Our adaptive BFS algorithm is based on performance prediction, which in turn is
based on a machine learning model. Performance prediction based on machine
learning has been attempted in many instances in the
past~\cite{Zhang2011,Song2013,Lee2007,Joseph2006,Wu15hpca,Madougou16ashes}.
However, applying machine learning for an adaptive, level-switching BFS
requires significant changes: features and variable selection, as well as the
collection and selection of training data are specialized for the challenges of
graph analytics. To the best of our knowledge, we are the first to have
attempted training and using such a model for improving BFS performance by
runtime switching.

In summary, our work is the first to employ a performance model based on
machine learning for building a generalized version of the direction optimized
BFS~\cite{beamer2013direction}.

\section{Conclusion}
\label{sec:conclusion}
With the increased availability of large, complex graphs and the high demand
for their analysis, high performance computing techniques become mandatory to
handle large scale graph processing. Among these techniques, the use of
massively parallel architectures like GPUs has been successful in the past:
both new algorithms and new processing systems have been proposed to speed-up
large scale graph processing.

Yet, despite the rapid innovation in the field, there has been little progress
in quantifying the actual correlation  between graph properties and the
performance of graph analytics. In other words, the performance variability of
graph processing, visible for most algorithms when processing different graphs
\emph{and} even when processing different layers of the same graph, has not
been quantified and/or addressed.

In this work, we propose to use this variability to gain performance for a
given graph processing algorithm: BFS traversal on GPUs. Our approach works as
follows: given a set of BFS algorithms (15 in total), and an input dataset, we
aim to determine \emph{and} employ, for \emph{each level} in the BFS traversal,
the best algorithm in the available set. This is an generalization of the work
on direction-switching BFS~\cite{beamer2013direction} and adaptive graph
algorithms~\cite{becchi2013ipdps}, to which we have added a much more
systematic switching detection.

Specifically, we use machine learning concepts to train a prediction model,
which is used at runtime to determine if switching is needed and, if so, to
which variant we should switch. This combination of machine learning modeling
and the large set of algorithms makes our approach competitive with
state-of-the-art graph processing systems and algorithms.

Our findings are interesting in two different ways. First, we demonstrate high
performance, with an average gain of $2{\times}$ over Gunrock and $15{\times}$
over LonestarGPU. Second, and more relevant for the original contribution of
this work, we demonstrate that machine learning can be used to build a
high-accuracy model which, taking into account graph and algorithm properties,
can predict the optimal selection of BFS implementations for ${\sim}56\%$ of
all BFS traversals and within $2{\times}$ of optimal for ${\sim}97\%$ of all
traversals.

We conclude that this work is a step forward in quantifying and using the
impact of graph properties on the performance of graph processing. For the near
future, we plan to pursue three research objectives: design and automate an
efficient training process, investigate the potential contribution of other BFS
algorithms, and test other modeling techniques that offer a good balance
between accuracy and speed of run-time evaluation. Finally, on the long term,
we plan to use these results to actually understand the impact of graph
properties on BFS graph traversal on GPUs.

\bibliographystyle{IEEEtranN}
{\footnotesize
\bibliography{main}}

\begin{thebibliography}{46}
\providecommand{\natexlab}[1]{#1}
\providecommand{\url}[1]{#1}
\csname url@samestyle\endcsname
\providecommand{\newblock}{\relax}
\providecommand{\bibinfo}[2]{#2}
\providecommand{\BIBentrySTDinterwordspacing}{\spaceskip=0pt\relax}
\providecommand{\BIBentryALTinterwordstretchfactor}{4}
\providecommand{\BIBentryALTinterwordspacing}{\spaceskip=\fontdimen2\font plus
\BIBentryALTinterwordstretchfactor\fontdimen3\font minus
  \fontdimen4\font\relax}
\providecommand{\BIBforeignlanguage}[2]{{%
\expandafter\ifx\csname l@#1\endcsname\relax
\typeout{** WARNING: IEEEtranN.bst: No hyphenation pattern has been}%
\typeout{** loaded for the language `#1'. Using the pattern for}%
\typeout{** the default language instead.}%
\else
\language=\csname l@#1\endcsname
\fi
#2}}
\providecommand{\BIBdecl}{\relax}
\BIBdecl

\bibitem[Avery(2011)]{avery2011giraph}
C.~Avery, ``Giraph: Large-scale graph processing infrastructure on hadoop,''
  \emph{Proceedings of the Hadoop Summit. Santa Clara}, 2011.

\bibitem[Hong et~al.(2015)Hong, Depner, Manhardt, Van Der~Lugt, Verstraaten,
  and Chafi]{hong2015pgx}
S.~Hong, S.~Depner, T.~Manhardt, J.~Van Der~Lugt, M.~Verstraaten, and H.~Chafi,
  ``{PGX.D}: A fast distributed graph processing engine,'' in \emph{Proceedings
  of the International Conference for High Performance Computing, Networking,
  Storage and Analysis}.\hskip 1em plus 0.5em minus 0.4em\relax ACM, 2015,
  p.~58.

\bibitem[Guo et~al.(2015{\natexlab{a}})Guo, Varbanescu, Iosup, and
  Epema]{YGuo2015ccgrid}
Y.~Guo, A.~L. Varbanescu, A.~Iosup, and D.~Epema, ``{An Empirical Performance
  Evaluation of GPU-Enabled Graph-Processing Systems},'' in \emph{CCGrid'15},
  2015.

\bibitem[Lu et~al.(2014)Lu, Cheng, Yan, and Wu]{journal/VLDB/LuCYW14}
Y.~Lu, J.~Cheng, D.~Yan, and H.~Wu, ``{Large-Scale Distributed Graph Computing
  Systems: An Experimental Evaluation},'' \emph{VLDB}, 2014.

\bibitem[Han et~al.(2014)Han, Daudjee, Ammar, Ozsu, Wang, and
  Jin]{journal/VLDB/HanDAOWJ14}
M.~Han, K.~Daudjee, K.~Ammar, M.~T. Ozsu, X.~Wang, and T.~Jin, ``{An
  Experimental Comparison of Pregel-Like Graph Processing Systems},''
  \emph{VLDB}, 2014.

\bibitem[Elser and Montresor(2013)]{conf/bigdata/ElserM13}
B.~Elser and A.~Montresor, ``{An Evaluation Study of Bigdata Frameworks for
  Graph Processing},'' in \emph{Big Data}, 2013.

\bibitem[Satish et~al.(2014)Satish, Sundaram, Patwary, Seo, Park, Hassaan,
  Sengupta, Yin, and Dubey]{conf/SIGMOD/SatishSPSPHSYD14}
N.~Satish, N.~Sundaram, M.~A. Patwary, J.~Seo, J.~Park, M.~A. Hassaan,
  S.~Sengupta, Z.~Yin, and P.~Dubey, ``{Navigating the Maze of Graph Analytics
  Frameworks using Massive Graph Datasets},'' in \emph{SIGMOD}, 2014.

\bibitem[Guo et~al.(2014)Guo, Biczak, Varbanescu, Iosup, Martella, and
  Willke]{conf/IPDPS/GuoBVIMW14}
Y.~Guo, M.~Biczak, A.~L. Varbanescu, A.~Iosup, C.~Martella, and T.~L. Willke,
  ``{How Well do Graph-Processing Platforms Perform? An Empirical Performance
  Evaluation and Analysis},'' in \emph{IPDPS}, 2014.

\bibitem[Committee(2010)]{graph500}
T.~G. .~S. Committee. (2010) The graph 500 list.

\bibitem[Beamer et~al.(2013)Beamer, Asanovi{\'c}, and
  Patterson]{beamer2013direction}
S.~Beamer, K.~Asanovi{\'c}, and D.~Patterson, ``Direction-optimizing
  breadth-first search,'' \emph{Scientific Programming}, vol.~21, no. 3-4, pp.
  137--148, 2013.

\bibitem[Bulu\c{c} et~al.(2016 (in press))Bulu\c{c}, Beamer, Madduri,
  Asanovi\'{c}, and Patterson]{Buluc16graph}
\BIBentryALTinterwordspacing
A.~Bulu\c{c}, S.~Beamer, K.~Madduri, K.~Asanovi\'{c}, and D.~Patterson,
  ``Distributed-memory breadth-first search on massive graphs,'' in
  \emph{Parallel Graph Algorithms}, D.~Bader, Ed.\hskip 1em plus 0.5em minus
  0.4em\relax CRC Press, Taylor-Francis, 2016 (in press). [Online]. Available:
  \url{http://gauss.cs.ucsb.edu/~aydin/ChapterBFS2015.pdf}
\BIBentrySTDinterwordspacing

\bibitem[A.~Lumsdaine and Berry(2007)]{Lumsdaine2007}
B.~H. A.~Lumsdaine, D.~Gregor and J.~W. Berry, ``Challenges in parallel graph
  processing,'' \emph{Parallel Processing Letters 17}, 2007.

\bibitem[Wang et~al.(2016)Wang, Davidson, Pan, Wu, Riffel, and
  Owens]{wang2016gunrock}
Y.~Wang, A.~Davidson, Y.~Pan, Y.~Wu, A.~Riffel, and J.~D. Owens, ``Gunrock: A
  high-performance graph processing library on the gpu,'' in \emph{Proceedings
  of the 21st ACM SIGPLAN Symposium on Principles and Practice of Parallel
  Programming}.\hskip 1em plus 0.5em minus 0.4em\relax ACM, 2016, p.~11.

\bibitem[Heldens et~al.(2016)Heldens, Varbanescu, and Iosup]{Heldens16IA3}
S.~Heldens, A.~L. Varbanescu, and A.~Iosup, ``Dynamic load balancing for
  high-performance graph processing on hybrid cpu-gpu platforms,'' in
  \emph{2016 6th Workshop on Irregular Applications: Architecture and
  Algorithms (IA3)}, Nov 2016, pp. 62--65.

\bibitem[Khorasani et~al.(2014)Khorasani, Vora, Gupta, and
  Bhuyan]{khorasani2014cusha}
F.~Khorasani, K.~Vora, R.~Gupta, and L.~N. Bhuyan, ``{CuSha}: vertex-centric
  graph processing on {GPUs},'' in \emph{{HPCS}}.\hskip 1em plus 0.5em minus
  0.4em\relax ACM, 2014, pp. 239--252.

\bibitem[Li and Becchi(2013)]{becchi2013ipdps}
D.~Li and M.~Becchi, ``Deploying graph algorithms on gpus: An adaptive
  solution,'' in \emph{{IPDPS 2013}}, May 2013, pp. 1013--1024.

\bibitem[Bulu\c{c} et~al.(2010)Bulu\c{c}, Gilbert, and Budak]{gpuapsp:2008-15}
\BIBentryALTinterwordspacing
A.~Bulu\c{c}, J.~R. Gilbert, and C.~Budak, ``Solving path problems on the
  {GPU},'' \emph{Parallel Computing}, vol.~36, no. 5-6, pp. 241 -- 253, 2010.
  [Online]. Available:
  \url{http://gauss.cs.ucsb.edu/publication/parco_apsp.pdf}
\BIBentrySTDinterwordspacing

\bibitem[Gonzalez et~al.(2012)Gonzalez, Low, Gu, Bickson, and
  Guestrin]{gonzalez2012powergraph}
J.~E. Gonzalez, Y.~Low, H.~Gu, D.~Bickson, and C.~Guestrin, ``Powergraph:
  Distributed graph-parallel computation on natural graphs.'' in \emph{OSDI},
  vol.~12, no.~1, 2012, p.~2.

\bibitem[Xin et~al.(2013)Xin, Gonzalez, Franklin, and Stoica]{xin2013graphx}
R.~S. Xin, J.~E. Gonzalez, M.~J. Franklin, and I.~Stoica, ``Graphx: A resilient
  distributed graph system on spark,'' in \emph{First International Workshop on
  Graph Data Management Experiences and Systems}.\hskip 1em plus 0.5em minus
  0.4em\relax ACM, 2013, p.~2.

\bibitem[Malewicz et~al.(2010)Malewicz, Austern, Bik, Dehnert, Horn, Leiser,
  and Czajkowski]{malewicz2010pregel}
G.~Malewicz, M.~H. Austern, A.~J. Bik, J.~C. Dehnert, I.~Horn, N.~Leiser, and
  G.~Czajkowski, ``Pregel: a system for large-scale graph processing,'' in
  \emph{Proceedings of the 2010 ACM SIGMOD International Conference on
  Management of data}.\hskip 1em plus 0.5em minus 0.4em\relax ACM, 2010, pp.
  135--146.

\bibitem[McCune et~al.(2015)McCune, Weninger, and Madey]{McCune:2015:TLV}
R.~R. McCune, T.~Weninger, and G.~Madey, ``Thinking like a vertex: A survey of
  vertex-centric frameworks for large-scale distributed graph processing,''
  \emph{ACM Comput. Surv.}, vol.~48, no.~2, Oct 2015.

\bibitem[Varbanescu et~al.(2015)Varbanescu, Verstraaten, Penders, Sips, and
  de~Laat]{varbanescu15icpe}
A.~L. Varbanescu, M.~Verstraaten, A.~Penders, H.~Sips, and C.~de~Laat, ``{Can
  Portability Improve Performance? An Empirical Study of Parallel Graph
  Analytics},'' in \emph{ICPE'15}, 2015.

\bibitem[Verstraaten et~al.(2015)Verstraaten, Varbanescu, and
  de~Laat]{verstraaten2015quantifying}
M.~Verstraaten, A.~L. Varbanescu, and C.~de~Laat, ``Quantifying the performance
  impact of graph structure on neighbour iteration strategies for pagerank,''
  in \emph{Euro-Par 2015: Parallel Processing Workshops}.\hskip 1em plus 0.5em
  minus 0.4em\relax Springer, 2015, pp. 528--540.

\bibitem[Lehnert et~al.(2016)Lehnert, Berrendorf, Ecker, and
  Mannuss]{lehnert2016performance}
C.~Lehnert, R.~Berrendorf, J.~P. Ecker, and F.~Mannuss, ``Performance
  prediction and ranking of spmv kernels on gpu architectures,'' in
  \emph{European Conference on Parallel Processing}.\hskip 1em plus 0.5em minus
  0.4em\relax Springer, 2016, pp. 90--102.

\bibitem[Leskovec(2006)]{SNAP}
J.~Leskovec, ``{S}tanford {N}etwork {A}nalysis {P}latform ({SNAP}),''
  \emph{Stanford University}, 2006.

\bibitem[Kunegis(2013)]{KONECT}
J.~Kunegis, ``Konect: The koblenz network collection,'' in \emph{Proceedings of
  the 22Nd International Conference on World Wide Web}, ser. WWW '13 Companion,
  2013, pp. 1343--1350.

\bibitem[Chakrabarti et~al.(2004)Chakrabarti, Zhan, and
  Faloutsos]{chakrabarti2004r}
D.~Chakrabarti, Y.~Zhan, and C.~Faloutsos, ``R-mat: A recursive model for graph
  mining.'' in \emph{SDM}, vol.~4.\hskip 1em plus 0.5em minus 0.4em\relax SIAM,
  2004, pp. 442--446.

\bibitem[Leskovec et~al.(2010)Leskovec, Chakrabarti, Kleinberg, Faloutsos, and
  Ghahramani]{leskovec2010kronecker}
J.~Leskovec, D.~Chakrabarti, J.~Kleinberg, C.~Faloutsos, and Z.~Ghahramani,
  ``Kronecker graphs: An approach to modeling networks,'' \emph{The Journal of
  Machine Learning Research}, vol.~11, pp. 985--1042, 2010.

\bibitem[Holme and Kim(2002)]{holme2002growing}
P.~Holme and B.~J. Kim, ``Growing scale-free networks with tunable
  clustering,'' \emph{Physical review E}, vol.~65, no.~2, p. 026107, 2002.

\bibitem[Verstraaten et~al.(2016)Verstraaten, Varbanescu, and
  de~Laat]{verstraaten2016}
M.~Verstraaten, A.~L. Varbanescu, and C.~de~Laat, ``Synthetic graph generation
  for systematic exploration of graph structural properties,'' in
  \emph{Euro-Par 2016: Parallel Processing Workshops}.\hskip 1em plus 0.5em
  minus 0.4em\relax Springer, 2016.

\bibitem[Burtscher et~al.(2012)Burtscher, Nasre, and
  Pingali]{burtscher2012quantitative}
M.~Burtscher, R.~Nasre, and K.~Pingali, ``A quantitative study of irregular
  programs on gpus,'' in \emph{Workload Characterization (IISWC), 2012 IEEE
  International Symposium on}.\hskip 1em plus 0.5em minus 0.4em\relax IEEE,
  2012, pp. 141--151.

\bibitem[Breiman et~al.(1984)Breiman, Friedman, Stone, and
  Olshen]{breiman1984classification}
L.~Breiman, J.~Friedman, C.~J. Stone, and R.~A. Olshen, \emph{Classification
  and Regression Trees}.\hskip 1em plus 0.5em minus 0.4em\relax CRC press,
  1984.

\bibitem[Pedregosa et~al.(2011)Pedregosa, Varoquaux, Gramfort, Michel, Thirion,
  Grisel, Blondel, Prettenhofer, Weiss, Dubourg, Vanderplas, Passos,
  Cournapeau, Brucher, Perrot, and Duchesnay]{scikit-learn}
F.~Pedregosa, G.~Varoquaux, A.~Gramfort, V.~Michel, B.~Thirion, O.~Grisel,
  M.~Blondel, P.~Prettenhofer, R.~Weiss, V.~Dubourg, J.~Vanderplas, A.~Passos,
  D.~Cournapeau, M.~Brucher, M.~Perrot, and E.~Duchesnay, ``Scikit-learn:
  Machine learning in {P}ython,'' \emph{Journal of Machine Learning Research},
  vol.~12, pp. 2825--2830, 2011.

\bibitem[Volkov(2016)]{volkov2016understanding}
V.~Volkov, ``Understanding latency hiding on gpus,'' 2016.

\bibitem[NVIDIA()]{nvidiacuda}
NVIDIA, ``Cuda toolkit documentation v8.0.61,''
  \url{http://docs.nvidia.com/cuda/index.html#axzz4dqjRLhKW}.

\bibitem[S.~Hong and Olukotun(2011)]{Hong2011}
T.~O. S.~Hong, S. K.~Kim and K.~Olukotun, ``Accelerating cuda graph algorithms
  at maximum warp,'' \emph{Principles and Practice of Parallel Programming,
  PPoPP'11}, 2011.

\bibitem[Harris(2007)]{harris2007optimizing}
M.~Harris, ``Optimizing cuda,'' \emph{SC07: High Performance Computing With
  CUDA}, 2007.

\bibitem[Bulu\c{c} et~al.(2011)Bulu\c{c}, Gilbert, and Shah]{sparsechapter}
A.~Bulu\c{c}, J.~R. Gilbert, and V.~B. Shah, ``{Implementing Sparse Matrices
  for Graph Algorithms},'' in \emph{Graph Algorithms in the Language of Linear
  Algebra}, J.~Kepner and J.~R. Gilbert, Eds.\hskip 1em plus 0.5em minus
  0.4em\relax SIAM Press, 2011.

\bibitem[Doekemeijer and Varbanescu(2014)]{Doekemeijer14survey}
\BIBentryALTinterwordspacing
N.~Doekemeijer and A.~L. Varbanescu, ``A survey of parallel graph processing
  frameworks,,'' TUDelft, Tech. Rep. PDS-2014-003, 2014. [Online]. Available:
  \url{http://www.ds.ewi.tudelft.nl/fileadmin/pds/reports/2014/PDS-2014-003.pdf}
\BIBentrySTDinterwordspacing

\bibitem[Guo et~al.(2015{\natexlab{b}})Guo, Varbanescu, Iosup, and
  Epema]{guo2015empirical}
Y.~Guo, A.~L. Varbanescu, A.~Iosup, and D.~Epema, ``An empirical performance
  evaluation of gpu-enabled graph-processing systems,'' in \emph{Cluster, Cloud
  and Grid Computing (CCGrid), 2015 15th IEEE/ACM International Symposium
  on}.\hskip 1em plus 0.5em minus 0.4em\relax IEEE, 2015, pp. 423--432.

\bibitem[Zhang et~al.(2011)Zhang, Hu, Li, and Peng]{Zhang2011}
Y.~Zhang, Y.~Hu, B.~Li, and L.~Peng, ``Performance and power analysis of {ATI
  GPU}: A statistical approach,'' in \emph{NAS'11}.\hskip 1em plus 0.5em minus
  0.4em\relax IEEE Computer Society, 2011.

\bibitem[Song et~al.(2013)Song, Su, Rountree, and Cameron]{Song2013}
S.~Song, C.~Su, B.~Rountree, and K.~W. Cameron, ``A simplified and accurate
  model of power-performance efficiency on emergent {GPU} architectures,'' in
  \emph{IPDPS '13}.\hskip 1em plus 0.5em minus 0.4em\relax IEEE Computer
  Society, 2013.

\bibitem[Lee et~al.(2007)Lee, Brooks, de~Supinski, Schulz, Singh, and
  McKee]{Lee2007}
\BIBentryALTinterwordspacing
B.~C. Lee, D.~M. Brooks, B.~R. de~Supinski, M.~Schulz, K.~Singh, and S.~A.
  McKee, ``Methods of inference and learning for performance modeling of
  parallel applications,'' in \emph{Proceedings of the 12th ACM SIGPLAN
  Symposium on Principles and Practice of Parallel Programming}, ser. PPoPP
  '07.\hskip 1em plus 0.5em minus 0.4em\relax New York, NY, USA: ACM, 2007, pp.
  249--258. [Online]. Available:
  \url{http://doi.acm.org/10.1145/1229428.1229479}
\BIBentrySTDinterwordspacing

\bibitem[Joseph et~al.(2006)Joseph, Vaswani, and Thazhuthaveetil]{Joseph2006}
P.~J. Joseph, K.~Vaswani, and M.~J. Thazhuthaveetil, ``{Construction and use of
  linear regression models for processor performance analysis},'' in
  \emph{International Symposium on High-Performance Computer Architecture},
  2006, pp. 99--108.

\bibitem[Wu et~al.(2015)Wu, Greathouse, Lyashevsky, Jayasena, and
  Chiou]{Wu15hpca}
G.~Wu, J.~Greathouse, A.~Lyashevsky, N.~Jayasena, and D.~Chiou, ``Gpgpu
  performance and power estimation using machine learning,'' in \emph{High
  Performance Computer Architecture (HPCA), 2015 IEEE 21st International
  Symposium on}, Feb 2015, pp. 564--576.

\bibitem[Madougou et~al.(2016)Madougou, Varbanescu, Laat, and
  Nieuwpoort]{Madougou16ashes}
S.~Madougou, A.~L. Varbanescu, C.~D. Laat, and R.~V. Nieuwpoort, ``A tool for
  bottleneck analysis and performance prediction for gpu-accelerated
  applications,'' in \emph{2016 IEEE International Parallel and Distributed
  Processing Symposium Workshops (IPDPSW)}, May 2016, pp. 641--652.

\end{thebibliography}
\end{document}